\documentclass[12pt, epsfig]{article}
\usepackage{amsmath,amsfonts,amssymb,amsthm,amstext,amscd,eucal,graphicx , xcolor}

\usepackage[all]{xy}
\usepackage{dsfont}
\usepackage{hyperref}
\usepackage{amsmath}
\usepackage{caption}
\usepackage{cite}

\makeatletter \@addtoreset{equation}{section}

\makeatletter\renewcommand\section{\@startsection {section}{1}{\z@}%
                                   {-3.5ex \@plus -1ex \@minus -.2ex}
                                   {2.3ex \@plus.2ex}%
                                   {\normalfont\large\bfseries}}
\renewcommand\subsection{\@startsection{subsection}{2}{\z@}%
                                     {-3.25ex\@plus -1ex \@minus -.2ex}%
                                     {1.5ex \@plus .2ex}%
                                     {\normalfont\bfseries}}

\newcommand{\be}{\begin{equation}}
\newcommand{\ee}{\end{equation}}
\newcommand{\bea}{\begin{eqnarray}}
\newcommand{\eea}{\end{eqnarray}}
\newcommand{\bse}{\begin{subequations}}
\newcommand{\ese}{\end{subequations}}
\newcommand{\beqa}{\begin{eqnarray}}
\newcommand{\eeqa}{\end{eqnarray}}
\newcommand{\beqar}{\begin{eqnarray*}}
\newcommand{\eeqar}{\end{eqnarray*}}
\newcommand{\bi}{\begin{itemize}}
\newcommand{\ei}{\end{itemize}}
\newcommand{\bn}{\begin{enumerate}}
\newcommand{\en}{\end{enumerate}}

\newcommand{\ba}{\begin{array}}
\newcommand{\ea}{\end{array}}
\newcommand{\bc}{\begin{center}}
\newcommand{\ec}{\end{center}}

\newcommand{\etal}{{\em et al.}\ }

\newcommand{\xdownarrow}[1]{%
	{\left\downarrow\vbox to #1{}\right.\kern-\nulldelimiterspace}}

\newcommand{\eps}{\varepsilon}

\definecolor{darkgreen}{rgb}{0,0.3,0}
\definecolor{darkblue}{rgb}{0,0,0.3}
\definecolor{darkred}{rgb}{0.7,0,0}

\newcommand{\old}[1]{}

\begin{document}

\begin{titlepage}

\begin{flushright}\vspace{-3cm}
{
\today }\end{flushright}
\vspace{-.5cm}

\begin{center}

\large{\bf{Higher Dimensional Charged AdS Black Holes at Ultra-spinning Limit\\ and Their 2d CFT Duals}}

\bigskip\bigskip

\large{\bf{S.M. Noorbakhsh\footnote{e-mail:
m.noorbakhsh@semnan.ac.ir}$^{ }$
and  M. Ghominejad\footnote{e-mail: mghominejad@semnan.ac.ir }$^{}$  }}
\\

\vspace{5mm}
\normalsize
\bigskip\medskip
{\it Department of Physics, Semnan University, P.O. Box 35195-363, Semnan, Iran}
\smallskip
\\

\end{center}
\setcounter{footnote}{0}

\begin{abstract}
\noindent
By using the ultra-spinning limit as a generating solution technique, we construct a novel class of charged rotating asymptotic AdS black holes. That describes the exact D-dimnsioanl solutions of Einstein-Maxwell dilaton theory in the presence of negative cosmological constant. The obtained geometries possess some punctures, describing a noncompact horizon, but has a finite area. We then explicitly investigate the validity of the Kerr/CFT correspondence for all dimensional cases. We find a main result for the central charges associated to $[(d-1)/2]$ copies of dual $2D$ CFTs. We then argue the existence of a precise agreement between the microscopic entropy of dual CFTs and the entropy of the noncompact horizon.

\end{abstract}

\end{titlepage}
\renewcommand{\baselinestretch}{1.1}  

\section{Introduction} 
      
It was shown that within the higher-dimensional general relativity one can access to the rotating black holes with any arbitrary large angular momentum by keeping the mass fixed \cite{MyersPerry:1986}. In order to study the stability of Myers-Perry black hole in a large angular momentum per unit mass, such that the rotation parameter $a \rightarrow \infty$, Emparan and Myers introduced the "ultra-spinning" black holes \cite{EmparanMyers:2003}. The horizon geometry of their resulting solutions have a “pancake” shape, spreading out along the rotation plane while are contracted in the transverse directions. They also found a limit in which these rapid black holes become an unstable black membrane at a finite large rotation. It was generalized an analogous analysis for asymptotic AdS rotating black holes. In $d \ge 6$ what is called as a black brane ultra-spinning limit proposed in \cite{CaldarelliEmparan:2008} by which the rotation parameter $a_i$ approaches to the AdS radius $l$, while one has to keep the physical mass and angular momentum finite, thereby leads to a static black brane. Furthermore, a different technique was introduced in \cite{ArmasObers} that ultra-spinning (A)dS solutions can be derived such that both $a$ and $l$ simultaneously approach to infinity while imposing $a/l$ fixed. Meanwhile,  Caldarelli \etal studied other applicable technique  \cite{CaldarelliEmparan:2008,CaldarelliLeigh:2011},
in which by taking $a \rightarrow l$ limit the horizon radius $r_+$ remains fixed, while the polar angle $\theta$ is scaled in a special way to avoid a metric singularity. This limit gives rise to a noncompactness black hole solution with a horizon topology as $\mathbb{H}^2 \times S^{d-4} $. So this ultra-spinning method refers to a hyperboloid membrane limit.

In \cite{HennigarKubiznakMann:2014} it was introduced a novel ultra-spinning approach that can be used only in the presence of the negative cosmological constant. The constructing geometries have a noncompact horizon that topologically can be described as a sphere with some punctures, but has a finite area. These class of black holes, in a sense, correspond to those were constructed in $\mathcal{N}=2$ gauged supergravity in the presence of a scalar potential, which have similar horizon topology \cite{GneccchiHristovKlemm:2014, Klemm:2014}

In the ultra-spinning technique introduced \cite{HennigarKubiznakMann:2014} to gain a regular metric, one should first transform the given rotating AdS black hole to an asymptotically rotating frame, then before boosting this rotation to the speed of light by taking $a \rightarrow l$, one should replace the corresponding azimuthal coordinate by a new scaled proper coordinate. This method has been investigated for the general multi-spinning Kerr-AdS black holes in \cite{HennigarKubiznakMann:2015}, and for two particular classes of gauged supergravity solutions in \cite{Noorbakhsh:2016}. Recently this ultra-spinning limit is developed to a class of extremal vanishing horizon in odd dimensions \cite{Noorbakhsh:201708}. The obtained geometries as a new classes of ultra-spinning solutions exhibit a distinguishable property in the extended thermodynamic phase space point of view, where the variation of the cosmological constant has to be taken in to account \cite{KastorRayTraschen:2009}. It was shown that the entropy of these specific solutions in some range of parameter space violate the Reverse Isoperimetric Inequality. So, this ultra-spinning limit is denoted as super-entropic limit as well. \cite{HennigarKubiznakMann:2014,HennigarKubiznakMann:2015}.\footnote[1]{It was shown that for a large class of black holes, for a given thermodynamic volume the maximum entropy is related to the (charged) AdS-Schwarzschild black hole. The `superentropic' black holes referring to those their entropy exceed the maximal entropy \cite{CveticGibbonsKubiznak-ISO:2012}.} 

There are strong evidences indicating black
holes can be viewed as thermodynamic systems possessing underlying microstate structure, attributed to a theory of quantum gravity. To provide an appropriate quantum gravity theory, proposing a successful formalism to calculate the entropy of black holes is one of the most crucial challenge. It was a long historical idea supporting that only microstates near the black hole horizon should be considered to calculate the black hole entropy \cite{NH1,NH10,NH11,NH12}. For extremal black holes, the Kerr/CFT correspondence gives a remarkable observation that the statistical microscopic entropy of the quantum gravity theory exactly matches with the Bekenstein-Hawking entropy of the black holes \cite{GuicaHartmanStrominger-KerrCFT}. The basic idea follows from the fundamental example of Brown and Henneaux \cite{BrownHenneaux:1986}, shown by considering consistent boundary conditions, the asymptotic symmetry group (ASG) of AdS$_3$ can be generated by two copies of Virasoro algebra. It asserts that the quantum theory of gravity on AdS$_3$ can be described by a two dimensional conformal field theory. Furthermore, Strominger \etal via the Cardy’s formula \cite{Cardy} found a precise agreement between the Bekenstein-Hawking entropy of the Kerr black holes and the entropy of the near-horizon quantum states \cite{NH10,HartmanStrominger:2009CFTDual}. Therefore what is dubbed the Kerr/CFT correspondence, conjectures an extreme Kerr black hole is holographically dual to a two-dimensional chiral (left-moving part) conformal field theory. These proposal have been developed to the large classes of higher dimensional black hole solutions in asymptotically flat and AdS spacetime  \cite{LuMeiPop-CFT:2008, ChowCvetic-CFT:2008, Compere:2009}. The near horizon geometry of an extremal $d$ dimensional Kerr black holes contain $[(d-1)/2]$ independent $U(1)$ isometry group, that by imposing a consistent boundary conditions they enhanced to $[(d-1)/2]$ copies of commutating Virasoro algebra, supporting the same numbers of CFTs dual. Therefore using the Cardy's formula one can find the microscopic entropy of all dual CFTs. Kerr/CFT correspondence predict the equality between the entropy of dual CFTs and macroscopic entropy of the black hole.  

There are numerous progresses to explore the validity of the Kerr/CFT within the various horizon topology. For instance, for the ultra-spinning solutions with a non-compact horizon, the existence of a well-defined Kerr/CFT is also confirmed by \cite{Noorbakhsh:2016} and \cite{Mann-SEBH-CFT:2015}.

The remarkable properties of ultra-spinning solutions, motivated us to further explore this limit onto a class of multi spinning charged AdS black holes in all higher dimensions. The main purpose of this work beside the better understanding the physics of charged AdS black holes in large angular momentum followed to generate a new class of charged AdS black hole solutions using the ultra-spinning technique. The black hole metric that we here consider describes multi-spinning solution of Einstein-Maxwell-Dilaton AdS black holes in all higher dimension was constructed by Wu \cite{KK-AdS:Wu:2011}. Since This class of solution obtained by using the Kaluza-Klein reduction method, they also refer to Kaluza-Klein charged AdS (KK-AdS) black holes. The first derivation of four dimensional charged KK-AdS black hole constructed in \cite{KK4d:1987} by a dimensional reduction of the boosted five dimensional neutral Kerr-AdS black hole, that presents an exact solution of four dimensional Einstein-Maxwell-Dilaton theory. Its generalization by including NUT charge obtained in \cite{KK4dNUT:2008}. A singly rotating general solutions in all higher dimensions were also found in \cite{KKd:2006}. In the context of gauged supergravity, the charged KK black hole solutions in four and five dimensions were introduced in  \cite{KKgauged1:2005,KKgauged2:2005} as well. The general  metric of multi-spinning charged KK-AdS black holes in all higher dimensions presented by Wu \cite{KK-AdS:Wu:2011}, that also describes an exact solution of $D$ dimensional Einstein-Maxwell dilaton theory.

In this work, we apply the novel ultra-spinning (super-entropic) limit proposing by \cite{HennigarKubiznakMann:2014}, onto the general multi-spinning charged KK-AdS black holes \cite{KK-AdS:Wu:2011}. Thereby we construct a new different class of higher dimensional charged rotating asymptotic AdS black hole solutions in that same theory.
Then we show the obtained geometries enjoy a noncompact horizon but has a finite area. Then we check the validity of the Kerr/CFT correspondence for our resulting solutions in all dimensions. Followed by finding the central charges associated to CFT duals and computing the microscopic entropy via the Cardy formula. 

The rest of this paper is organized as follows. In section \ref{GKK-AdS} we introduce the general charged KK-AdS black holes we shall analyze, reviewing their charges and thermodynamic. Then we obtain their general ultra-spinning version in all dimensions, followed by their horizon geometries showing a noncompact manifold. We then present the explicit metric for four dimensional case. In the Sec. \ref{SecKerrCFT} we study the ultra-spinning Kerr/CFT description, that in four dimensional case we precisely show that the entropy of the CFT side via the Cardy's formula agrees with the black hole entropy. Then in the last subsection, we find the near horizon geometry of our extremal ultra-spinning solutions in all dimensions. Showing that contain an AdS$_2$ sector product to a $S^{d-2}$ manifold having punctures. We also provide a main result confirming these non-compactness horizon solutions exhibit a well-defined Kerr/CFT correspondence.

\section{General charged rotating AdS black holes}\label{GKK-AdS}
Here, we consider the particular class of charged multi-spinning black holes as a solution of the Einstein-Maxwell dilaton theory 

\begin{equation}\label{ActionKK}
\mathcal{L} =\sqrt{-g}\big\{R -\frac{1}{4}(D-1)(D-2)(\partial \Phi)^2-\frac{1}{4}e^{-(D-1)\Phi}F^2 + g^2(D-1)[(D-3)e^{\Phi}+e^{-(D-3)\Phi}] \big\}.
\end{equation}
The general solution of this theory in all higher dimension using the Kaluza-Klien reduction technique is generated by Wu \cite{KK-AdS:Wu:2011} as
\begin{eqnarray}\label{Metric1}
ds^2&=&H^{\frac{1}{D-2}}\bigg[d\gamma^2+ \frac{U dr^2}{\Delta} + \frac{2m}{U H}\omega^2 + d\Omega^2\bigg],
\end{eqnarray}
where we have defined
\begin{eqnarray}  
d\gamma^2&=&-\frac{W \rho^2}{l^2} dt^2 + \sum_{i=1}^{N} \frac{r^2 + a_i^2}{\Xi_i}\mu_i^2 d\phi_i^2, \\ \nonumber
d\Omega^2&=&\sum_{i=1}^{N+\eps}\frac{r^2 + a_i^2}{\Xi_i}d\mu_i^2 -\frac{1}{ W \rho^2}\bigg(\sum_{i=1}^{N+\eps}\frac{r^2+a_i^2}{\Xi}\mu_i d\mu_i\bigg)^2, \\ \nonumber
\omega&=&c\, W dt-\sum_{i=1}^{N}\frac{a_i \sqrt{\chi_i}}{\Xi_i}\mu_i^2 d\phi\, ,
\end{eqnarray}
and 
\begin{eqnarray}\label{Metric2}
\rho^2&=&r^2+l^2, \qquad \qquad H=1+\frac{2ms^2}{U}, \quad \qquad U=r^\eps \sum_{i=1}^{N+\eps}\frac{\mu_i^2}{r^2+a_i^2}\prod_{j=1}^{N}(r^2+a_j^2), \qquad \qquad  \\ \nonumber
W&=&\sum_{i=1}^{N+\eps}\frac{\mu_i^2}{\Xi_i}, \qquad \qquad F=\frac{l^2r^2}{\rho^2}\sum_{i=1}^{N+\eps}\frac{\mu_i^2}{r^2 + a_i^2}, \qquad \qquad
f(r)=c^2 -s^2\rho^2/l^2,\\ \nonumber
\Delta&=&\frac{r^{\eps-2}\rho^2}{l^2}\prod_{i=1}^{N}(r^2+a_i^2)-2m f(r), \qquad \qquad
\chi_i=c^2-s^2 \Xi_i, \\ \nonumber
c&=&\cosh \delta, \qquad \qquad s=\sinh \delta, \qquad \qquad \qquad \Xi_i=1-a_i^2/l^2.
\end{eqnarray}
It should be noticed that to consistent the above general solution for any dimension $D$, one needs to define the following conventions
\begin{equation}\label{dim}
D=2N+1+\eps,
\end{equation}
where one set $\epsilon=0$ and  $1$ for even and odd dimensions respectively, and requiring $a_{N+1}=0$ for even dimensions. Also $N=[(D-1)/2]$ denotes the number of azimuthal directions $\phi_i$ corresponding to $N$ independent rotation parameters $a_i$ with periodicity $2\pi$. Moreover, there are $[D/2]$ numbers of "direction cosines" $\mu_i$'s as the remaining spatial coordinates subject to the following constraint
\begin{equation}\label{cosines}
\sum_{i=1}^{N+\eps}\mu_i^2=1.
\end{equation}
The gauge and dilaton field are also given by
\begin{eqnarray}\label{gaugefiled}
A&=&\frac{2m s}{U\,H}\big(c\, W dt- \sum_{i=1}^{N}\frac{a_i\sqrt{\chi_i}}{\Xi_i}\mu_i^2 d\phi\big), \qquad \quad \Phi=\frac{-1}{D-2}\ln(H).
\end{eqnarray}
Note, the metric (\ref{Metric1}) is known as a charged KK-AdS black hole solution. Here is written in an asymptotically static frame (ASF). In the uncharged case $(\delta=0)$, this solution reduces to the general class of Kerr-AdS black holes presenting in \cite{Gibbons:KerrAdS1} and \cite{Gibbons:KerrAdS2}. Also in the case of vanishing cosmological constant, the metric (\ref{Metric1}) corresponds to those introduced in \cite{Kunz:2006}.  Particularly, the supergravity solutions by only one electric charge which are derived in \cite{Cvetic:1999} are equivalent with a nonrotatting case of the metric (\ref{Metric1}) in D$=4, 5, 7$ dimensions.
 
\textbf{Charges and Thermodynamics: }
The  metric (\ref{Metric1}) describes a black hole, if one can find a real root from the equation $\Delta=0$. The outer root $(r_+)$ describes a killing horizon is generated by the killing vector $
K=\partial_t+\sum_{i}^{N} \Omega_i \partial_{\phi_i}$.
where $\Omega_i$'s denote the angular velocity on the horizon in the ASF. 

For the metric (\ref{Metric1}) the Hawking temperature and entropy are given by
\begin{eqnarray}\label{TemEnt}
T=\frac{\sqrt{f(r_+)}[V^\prime(r_+)-2mf^\prime(r_+)]}{4\pi^2 r_+^{\eps-2}c \prod_{i=1}^{N}(r_+^2+a_i^2)},
 \qquad  \qquad S=\frac{\mathcal{V}_{D-2}\,m\,r_+\,c\,l^2\sqrt{f(r_+)}}{2\rho_+^2)\prod_{i=1}^{N}\Xi_i},
\end{eqnarray}
where $\mathcal{V}_{D-2}$ refers to the volume of the unit $(D-2)$-sphere as
\begin{equation}
\mathcal{V}_{D-2}=\frac{2\pi^{(D-1)/2}}{\Gamma[(D-1)/2]}.
\end{equation}
Also the angular velocity and the electric potential on the horizon are 
\begin{eqnarray}\label{Omega}
\Omega^H_i=\frac{\rho_+^2a_i\sqrt{\chi_i}}{c\,l^2(r_+^2+a_i^2)}, \qquad \qquad \Phi^H=\frac{s}{c\,l^2}\rho_+^2.
\end{eqnarray}
The Mass, angular momenta and electric charge of this family of black hole solutions are calculated in \cite{KK-AdS:Wu:2011}, as follows
\begin{eqnarray}\label{Charges}
M&=&\frac{\mathcal{V}_{D-2}\, m}{8\pi \prod_{j=1}^{N} \Xi_j}\bigg[c^2\bigg(\sum_{i=1}^{N}\frac{2}{\chi_i}+\eps-2\bigg)+1\bigg], \\ \nonumber
J_i&=&\frac{\mathcal{V}_{D-2}\, m a_i c \sqrt{\chi_i}}{4 \pi \Xi_i \prod_{j=1}^{N}\Xi_i}, \qquad \qquad Q=\frac{(D-3)\mathcal{V}_{D-2}\, m cs}{8\pi\prod_{j=1}^{N}\Xi_j}.
\end{eqnarray}
It is shown in \cite{KK-AdS:Wu:2011} that these conserved charges satisfy the first law of thermodynamics 
\begin{eqnarray}\label{Smarr}
dM=TdS+\sum_{i=1}^{N}\Omega_i dJ_i+\Phi dQ,
\end{eqnarray}

\subsection{Ultra-Spinning limit}
In what follows we shall generate a new class of ultra-spinning charged AdS black holes based on the novel ultra-spinning technique introduced in \cite{HennigarKubiznakMann:2014}. This ultra-spinning method can be utilized by following three steps, i) to achieve a non-singular metric, one should transform the given rotating AdS black hole to an asymptotically rotating frame. Then the corresponding azimuthal coordinate should be scaled by a new coordinate transformation. ii) boosting the selected rotation to the speed of light easily by taking the limit $a \rightarrow l$. iii) finally, compactifying the new azimuthal coordinate. 

At the beginning, towards producing an ultra-spinning black hole upon the metric (\ref{Metric1}), we choose  
a certain $\varphi_j$ coordinate as an ultra-spinning direction. Before applying next steps, we start by a trick used in \cite{HennigarKubiznakMann:2015} for separating the $\phi_j$ direction and taking the limit $a_j \rightarrow l$. Thus, the following important result can be derived
\begin{equation}
W \Xi_j \rightarrow \mu_j^2. 
\end{equation}
Then using this relation, $d\Omega^2$ becomes
\begin{equation}\label{domegas}
d\Omega^2\, \rightarrow\, d\Omega_s^2=\sum_{i \neq j}^{N+\eps}\frac{r^2 + a_i^2}{\Xi_i}d\mu_i^2 -2\frac{d\mu_j}{\mu_j}\bigg(\sum_{i\neq j}^{N}\frac{r^2+a_i^2}{\Xi_i}\mu_i d\mu_i\bigg)+\frac{d\mu_j^2}{\mu_j^2}\big(\rho^2 \hat{W}+l^2\mu_j^2\big),
\end{equation}
where $\hat{W}=\sum_{i \neq j}\frac{\mu_i^2}{\Xi_i}$. Now we are ready to perform the ultra-spinning limit. So in the first step we have to use the following coordinate transformation to gain an asymptotically rotating frame (ARF),
\begin{equation}\label{Newphi}
\phi_j=\phi_j^R+\frac{a_j}{l^2}t.
\end{equation}
so, we get
\begin{equation}
\omega=(c\hat{W}+c\frac{\mu_j^2}{\Xi_j}-\frac{a_j^2 \sqrt{\chi_i}}{l^2 \Xi_j}\mu_j^2)dt - \frac{a_j \mu_j^2\sqrt{\chi_i}}{\Xi_j}d\phi_j^R-\sum_{i \neq j} \frac{a_i \sqrt{\chi_i}}{\Xi_i}\mu_i^2 d\phi_i.
\end{equation}
Now, by replacing the new coordinate transformation  
\begin{equation}
\varphi_j=\frac{\phi_j^R}{\Xi_j},
\end{equation}
and upon boosting $a \rightarrow l$, we find the new  geometry as
\begin{eqnarray}\label{MetricUS1}
ds^2&=&\hat{H}^{\frac{1}{D-2}}\bigg[d\gamma_s^2+ \frac{\hat{U} dr^2}{\hat{\Delta}} + \frac{2m}{\hat{U} \hat{H}}\omega_s^2\bigg] + d\Omega_s^2,
\end{eqnarray}
where 
\begin{eqnarray}\label{MetricUS2}\nonumber
w_s&=&(c\hat{W}+\frac{2+3s^2}{2c}\mu_j^2)dt-c\,l\mu_j^2 d\varphi_j-\sum_{i\neq j}^{N}\frac{a_i \mu_i^2 d\phi_i}{\Xi_i}, \\ \nonumber
d\gamma_s^2&=&-\bigg(\rho^2(\hat{W}+\mu_j^2)+\mu_j^2 l^2\bigg)\frac{dt^2}{l^2}+\rho^2 \mu_j^2 d\varphi_j^2+\frac{2\rho^2 \mu_j^2 dt d\varphi}{l}+ \sum_{i \neq j}\frac{r^2+a_i^2}{\Xi_i}\mu_i^2 d\phi_i^2,  \\ \nonumber
\hat{U}&=&r^\eps\bigg(\mu_j^2+\sum_{i\neq j} \frac{\mu_j^2 \rho^2}{r^2+a_j^2}\bigg)\prod_{k \neq j}^{N}(r^2+a_k^2),  \qquad \quad \hat{F}=\frac{r^2 l^2}{\rho^2}\bigg(\frac{\mu_j^2}{\rho^2}+\sum_{i \neq j}\frac{\mu_i^2}{r^2+a_i^2}\bigg), \\ 
\hat{\Delta}&=&\frac{r^{\eps-2}\rho^4}{l^2}\prod_{i\neq j}^{N}(r^2+a_i^2)-2m f(r). \qquad 
\end{eqnarray}
At the end, since $\varphi_j$ is a noncompact coordinate, one can identify it by requiring a periodicity such as
\begin{equation}\label{period}
\varphi_j \sim \varphi_j + \mu
\end{equation}

Also, the gauge field (\ref{gaugefiled}) under the ultra-spinning limit takes the following form
\begin{equation}\label{gaugeUS}
A=\frac{2m s}{\hat{U}\,\hat{H}}\bigg((c\hat{W}+\frac{2+3s^2}{2c}\mu_j^2)dt-c\,l\mu_j^2 d\varphi_j-\sum_{i\neq j}^{N}\frac{a_i \mu_i^2 d\phi_i}{\Xi_i}\bigg),
\end{equation}
We note that $\partial_{\varphi_j}$, is a Killing vector of obtained geometry (\ref{MetricUS1}). Therefore, straightforwardly one can show that the new metric is an exact solution of the theory (\ref{ActionKK}).  Moreover, the gauge field (\ref{gaugeUS}) satisfies the equation of motion of this theory as well. Also this new solution describe an asymptotic AdS geometry. The metric (\ref{Metric1}) in the case of vanishing charge $\delta=0$ reproduces the ultra-spinning general Kerr-AdS black holes investigating in \cite{HennigarKubiznakMann:2015}. 

\textbf{Basic properties: }
The new solution (\ref{MetricUS1}) describes an asymptotic AdS black hole whenever the equation $\hat{\Delta}=0$ allows a real root. The largest root $(r_+)$ can be generated by the Killing vector field
\begin{equation}\label{KillingVector}
K=\partial_t+\Omega_j \partial_{\varphi_j}+\Omega_i \partial_{\phi_i},
\end{equation}
where $\Omega_j$ and $\Omega_i$ are the angular velocities on the horizon 
\begin{equation}\label{AngularVelocity}
\Omega_{i\neq j}=\frac{(r_+^2+l^2)a_i\sqrt{\chi_i}}{c\,l^2(r_+^2+a_i^2)}, \qquad \qquad \Omega_j=\frac{l^2 (s^2+2)-r_0^2 s^2}{2 l (s^2+1) (l^2+r_0^2)}.
\end{equation}
Also for the new general solution  (\ref{MetricUS1}), the Hawking temperature can be derived through the
horizon surface gravity as
\begin{eqnarray}\label{TempUS}
T&=&\frac{\sqrt{f(r_+)}[\hat{V}^\prime(r_+)-2mf^\prime(r_+)]}{4\pi^2 \rho^2r_+^{\eps-2}c \prod_{i \neq j}^{N}(r_+^2+a_i^2)},
\end{eqnarray}
The electric potential on the horizon is calculated using the definition $\Phi^H=-K^\mu A_\mu|_{r=r_+}$, leading to
\begin{equation}\label{ElectricPotential}
\Phi^H=\frac{s\,(r_+^2+l^2)}{c\,l^2}.
\end{equation}
Let us here focus on the geometry of the horizon 
\begin{eqnarray}\label{Horizon}
ds^2_h&=&\hat{H}^{\frac{1}{D-2}}\bigg[\rho^2 \mu_j^2 d\varphi_j^2+ \sum_{i \neq j}\frac{r_+^2+a_i^2}{\Xi_i}\mu_i^2 d\phi_i^2 + \frac{2m}{\hat{U} \hat{H}} \bigg(c\,l\mu_j^2 d\varphi_j+\sum_{i\neq j}^{N}\frac{a_i \mu_i^2 d\phi_i}{\Xi_i}\bigg)\bigg] + d\Omega_s^2.
\end{eqnarray}
where the $d\Omega_s^2$ is given by \ref{domegas}. 
At the first glance to this part, it seems to be singular at $\mu_j=0$. But by examining the behavior of the horizon metric near $\mu_j=0$, one can show that these poles are not true curvature singularity. For simplicity we consider the $\phi_i=$const. and $\mu_i=$const. slices. Then the horizon metric \ref{Horizon} in the leading order expansion of small $\mu_j$ takes the following explanation
\begin{eqnarray}\label{Horizon-k}
ds^2_h&\approx&\hat{H}^{\frac{1}{D-2}}\rho^2 \hat{W}\bigg[\rho^2 \mu_j^2 d\varphi_j^2 + \frac{2 \,m}{\hat{W}\rho^2\hat{U} \hat{H}}c^2 l^2 \mu_j^4 d\varphi_j^2 + \frac{d\mu_j^2}{\mu_j^2} \bigg].
\end{eqnarray}
One can check that this metric present a constant negative curvature geometry on a quotient of  the hyperbolic space $\mathbb{H}^2$. Namely, the poles $\mu_j=0$ are no part of the spacetime but can be interpreted as a kind of boundary. Therefore any geometry of constant $(t,r)$ slices, and particularly the horizon metric describe noncompact manifolds, topologically are spheres containing some punctures. However one can find horizon area and entropy as a finite value 
\begin{equation}\label{Entropy}
S=\frac{\mathcal{V}_{D-2}\,r_+^{\epsilon-1}\,c\, }{4\sqrt{f(r_+)}}\prod_{i \neq j}^{N}\frac{r_+^2+a_i^2}{\Xi_i}.
\end{equation}

We should emphasis that it remains to be impossible to obtain a multi ultra-spinning solution. Because taking another coordinate as an ultra-spinning direction, leads the components of $d\Omega_s^2$ to diverge, which can not be removed by another new coordinate transformation similar to \ref{Newphi}

To better understanding of the obtained ultra-spinning geometry (\ref{MetricUS1}), we shall present the explicit metric for four dimensional case in the next subsections.

\subsection{Four Dimensional Ultra-Spinning  Charged AdS Black Holes}
In the four dimensional case, we have only one azimuthal coordinate $\phi$ corresponding to one rotational parameter $a$, and there is one spatial polar coordinate $\theta$. So, using the common convention
\begin{equation}\label{4dmu}
\mu_1=\sin\theta, \qquad \qquad \qquad \mu_2=\cos\theta,
\end{equation}
the explicit metric (\ref{MetricUS1}) in four dimension takes the following description
\begin{eqnarray}\label{ds4d} \nonumber
ds^2&=&\sqrt{1+\frac{2mrs^2}{\Delta_\theta}} \bigg[ \Delta_r dt^2 + \frac{l^2\,\Delta_\theta}{\Delta_r}dr^2 +\frac{2\,m r\,c^2}{\Delta_\theta+2mrs^2}\big(l\sin^2\theta d\varphi+\frac{\Sigma\, \cos^2\theta }{2m\,l^2\,c^2}dt \big)^2 
+ \frac{\Delta_\theta}{\sin^2\theta} d\theta^2
 \bigg],\\
\end{eqnarray}
where
\begin{eqnarray}\label{ds4d-2} \nonumber
\Delta_r&=&(r^2+l^2)^2-2mr(l^2-r^2s^2)}{2\,m\,r l^4 c^2, \qquad \qquad \Delta_\theta=r^2+l^2 \cos^2\theta,\\ 
\Sigma&=&\frac{l^2 (r^2+l^2+m r s^2)\cos^2\theta}{r}+2 m (r^2 s^2-l^2)+l^2(r-m s^2) +r^3.
\end{eqnarray}
For this black hole, the hawking temperature, entropy and angular velocity on the horizon read
\begin{eqnarray}\label{Thermo4d} \nonumber
T&=&\frac{\sqrt{l^2-r_+^2s^2}\,(l^4+r_+^4s^2-3c^2l^2r_+^2)}{4\pi\,c\,l^3r_+(l^2+r_+^2)}, \qquad \quad
S=\frac{\mu\,l\,c\,(l^2+r_+^2)}{2(l^2-r_+^2s^2)}.
\end{eqnarray}
In order to visualize the noncompactness horizon of the metric (\ref{ds4d}), one can embed the horizon metric in an Euclidean $3$-space. Using the procedure was taken in \cite{GneccchiHristovKlemm:2014, Noorbakhsh:2016}, we identify the horizon metric 
\begin{equation} \label{dsh}
ds_h^2=g_{\varphi\varphi}d\varphi^2+g_{\theta\theta}d\theta^2|_{r=r_+},
\end{equation}
with a line element in the cylindrical coordinates
\begin{equation} \label{CylindricalCo1}
 ds_3^2=dz^2+dR^2+R^2 d\Phi^2,
\end{equation}
 where $z = z(\theta)$, $R = R(\theta)$. Then using metric (\ref{ds4d}) and setting $\Phi=\frac{2\pi}{\mu} \varphi$, we have
 \begin{equation}\label{CylindricalCo2}
 R^2(\theta)=\bigg(\frac{\mu}{2\pi}\bigg)\,g_{\theta\theta},
 \end{equation}
 \begin{equation}\label{CylindricalCo3}\nonumber
 \bigg(\frac{dz(\theta)}{d\theta}\bigg)^2= \bigg(\frac{dR(\theta)}{d\theta}\bigg)^2-g_{\varphi\varphi}.
 \end{equation}
The surface of revolution by numerically integrating (\ref{CylindricalCo3}) is presented in Fig. 1. That clearly shows a manifold with two punctures at $\theta=0, \pi$.   
 \begin{figure}[htp]
 	\centering
 	\begin{tabular}{ccc}
 		\includegraphics[width=0.35\textwidth,height=0.35\textheight]{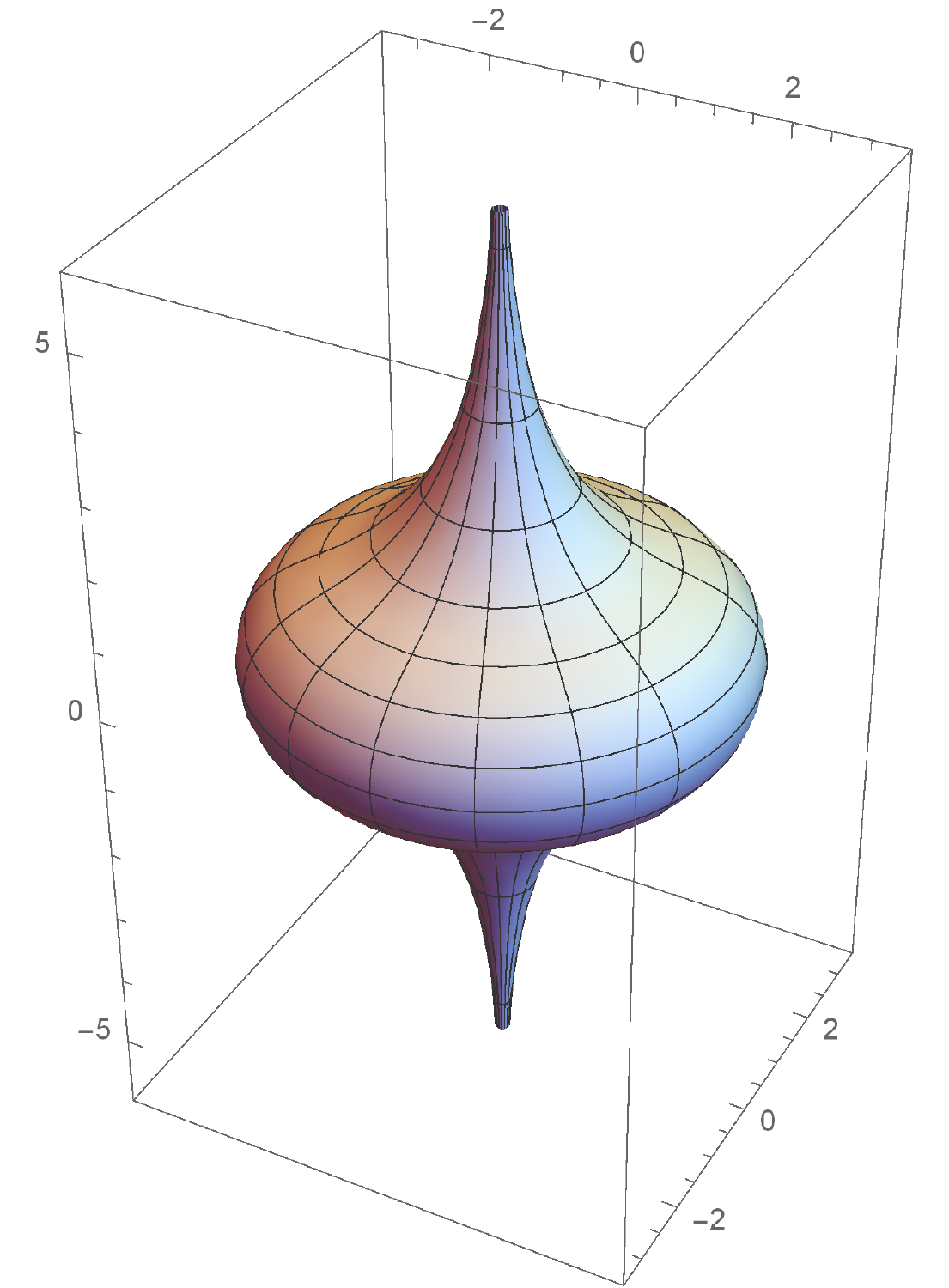}
 	\end{tabular}
 	\captionsetup{labelformat=empty}
 	\caption{{\bf Fig. 1. \bf Horizon embeddings in 4d}. Plot shows a $2$-dimensional horizons embedded in $\mathbb{R}$$^3$ as a surface of revolution, displaying topology of a sphere with two punctures. We have set $\mu = 2\pi, \, r_+ = 2,$ and $l=1$ with $s=0.4$ .  }\label{HorizonEmbedding4d}
 \end{figure}\label{Fig1-4dH}

\textbf{Conformal Boundary :}
Here, we find the conformal boundary of the metric (\ref{ds4d}) by taking conformal factor as  $l^2/r^2$, 
\begin{equation}\label{ConformalbdryUS4d}
 ds^2_{bdry}=-dt^2+2l\sin^2\theta\, dt d\varphi +\frac{l^2}{\sin^4\theta}d\theta^2.
\end{equation}
By introducing $k=(1-\cos\theta)$, this metric takes the following expression near the pole $\theta=0$ (small $k$)
\begin{equation}\label{Cbdry4dk}
ds^2_{bdry}=-dt^2+4lk\,dtd\varphi+\frac{l^2}{4k^2}.
\end{equation}
That shows an AdS$_3$ metric written as a Hopf-like fibration over $\mathbb{H}^2$, in which by approaching to the poles, $\varphi$ becomes a null coordinate. Therefore one can conclude that there is no problem near the poles $(\theta=0, \pi)$, they are not part of the spacetime indeed. It should be noted, one can study the behavior of any outgoing null geodesics that never reach to the $\theta=0, \pi$ axis in a finite affine parameter \cite{HennigarKubiznakMann:2015}. 

In the extended phase space thermodynamic it was shown in \cite{HennigarKubiznakMann:2014,HennigarKubiznakMann:2015} that the ultra-spinning general Kerr-AdS black hole solutions exhibit another distinguishable property. In which their entropy for some range of parameter space violate the `Reverse Isoperimetric Inequality', so they are called as super-entropic black holes. As a further work, we want to explore this approach for our new class of higher dimensional charged ultra-spinning solutions (\ref{MetricUS1}). 

\section{Kerr/CFT description}\label{SecKerrCFT}
In this section we explore the validity of the Kerr/CFT correspondence for the new obtained black holes (\ref{MetricUS1}). This duality that first proposed for $4$D extremal kerr black hole in \cite{GuicaHartmanStrominger-KerrCFT}, asserts this idea that whatever the concept of quantum gravity states in the near horizon region of an extremal Kerr black hole, they are dual to quantum states of a $2$D conformal field theory. The Virasoro algebra associated to dual CFT arises as an asymptotic symmetry group (ASG) of the near-horizon geometry. In fact one of the main ingredients of Kerr/CFT proposal is imposing a relevant boundary conditions leading to an ASG with a Virasoro algebra. 

In a $D$ dimensional multi-spinning extremal black hole one may find $[(D-1)/2]$ commuting Virasoro algebras supporting to same numbers of CFT duals. Namely there are $[(D-1)/2]$ independent central charges associated to each copy of dual CFTs. Then via the Cardy formula the microscopic entropy of all dual CFTs can be computed, in which they are exactly equal with each other and with the Bekenstein–
Hawking entropy of extremal black hole. It has been fully investigated in various dimensions \cite{LuMeiPop-CFT:2008} and different theory such as in gauged and ungauged supergravity solution \cite{ChowCvetic-CFT:2008}

Here, to start the procedure towards the Kerr/CFT correspondence for our ultra-spinning solution, we first consider the four dimensional metric (\ref{ds4d}). So, one need to find the near horizon geometry of the extremal version of the this metric. The extremality conditions obtained by imposing vanishing temperature ($T_H|_{r=r_0}=0$), as 
\begin{equation}\label{Ext4d}
m=\frac{(l^2+r_0^2)}{2r_0(r_0^2s^2-l^2)}, \qquad \qquad s=\frac{l}{r_0}\sqrt{\frac{l^2-3r_0^2}{3l^2-r_0^2}},
\end{equation}
where $r_0$ denotes the degenerate Killing horizon. Then we can expand the function $\hat{\Delta}$ near the horizon as
 \begin{eqnarray}\nonumber
 \hat{\Delta}=X(r-r_0)^2+\mathcal{O}(r-r_0)^3,
 \end{eqnarray}
 where
 \begin{eqnarray}
 X=\frac{3l^4-2l^2r_0^2+3r_0^4}{2l^2r_0^3}.
 \end{eqnarray}
The entropy of the metric (\ref{ds4d}) in the extremal limit reads
\begin{eqnarray}\label{EntExt}
S=\frac{\mu (l^2+r_0^2)}{2r_0}\sqrt{\frac{l^2-r_0^2}{2}}.
\end{eqnarray}

Now, to find the near horizon extremal geometry (NHEG) of the metric (\ref{ds4d}), one can employ the following dimensionless coordinate transformation 
\begin{equation}
r=r_0(1+\lambda \hat{r}), \qquad  \varphi=\hat{\varphi}+\Omega^0 \hat{t}, \qquad    \hat{t}=\frac{t}{2\pi s_0 r_0 \lambda}, \qquad \hat{\theta}=\theta, 
\end{equation}
where the scaling  $s_0=\frac{X}{l^2+r_0^2}\sqrt{\frac{2}{l^2-r_0^2}}$, and $\Omega^0=\Omega|_{r=r_0}$. Then, by taking $\lambda \rightarrow 0$, we obtain NHEG 
\begin{eqnarray}\label{ds4d-NH} 
ds^2&=&\sqrt{\hat{H}^0_4}\bigg[\frac{r_0^2+l^2\sin^2\theta}{X}\big(-\hat{r}^2 d\hat{t}^2 + \frac{d \hat{r}^2}{\hat{r}^2}\big) + \gamma(\theta)\bigg(d\varphi+k\hat{r}d\hat{t}\bigg)^2+F(\theta) d\theta^2
\bigg],
\end{eqnarray}
where 
\begin{eqnarray}\label{kds4}
\gamma(\theta)&=&\frac{(l^2+r_0^2)^2(l^2-r_0^2)\sin^4\theta}{l^4-r_0^4-2\,l^2r_0^2\sin^2\theta}, \qquad \quad F(\theta)= \frac{r_0^2+l^2 \cos^2\theta}{\sin^2\theta}, \\ \nonumber
\hat{H}^0_4&=&1+\frac{(l^2-3r_0^2)(l^2+r_0^2)}{2r_0^2(r_0^2+l^2\,\cos^2\theta)}\qquad \quad k=\frac{1}{2\pi T_L}=\frac{2\,l\,r_0}{s_0(l^2+r_0^2)^2}
\end{eqnarray}
where $T_l$ refer to the Frolov--Thorne temperature denotes the temperature associated to the left-moving part of the dual CFT.

The NHEG (\ref{ds4d-NH}) shows a direct product of AdS$_2\times U(1)$. As expected it contains the well-known AdS$_2$ sector, which is here written in Poincar$\acute{e}$-type coordinates. Appearing of the AdS$_2$ sector in the NHEG of any regular stationary black holes has been a great appeal in the literature \cite{KunduriLucietti, KunduriLucietti2, Astefanesei,JohnstoneSheikh-Jabbari}. It is shown that this geometry has the $SL(2, R)\times U(1)$ isometry group.

Also it was shown that this class of NHEG by imposing a set of consistent boundary conditions such as  \cite{GuicaHartmanStrominger-KerrCFT}
\begin{eqnarray}\label{ASG-Boundary}
\begin{pmatrix}
\mathcal{O}(r^2)& \mathcal{O}(1)& \mathcal{O}(1/r)&\mathcal{O}(1/r^2)\\

&\mathcal{O}(1)& \mathcal{O}(1/r)&\mathcal{O}(1/r)\\

& & \mathcal{O}(1/r)&\mathcal{O}(1/r^2)\\

& & &\mathcal{O}(1/r^3)\\
\end{pmatrix}.
\end{eqnarray}
in the coordinate of $(t,r,\phi,\theta)$, admits an enhanced Virasoro algebra \cite{GuicaHartmanStrominger-KerrCFT, Barnich}. Therefore, one can found an algebra of the charges associated to the asymptotic symmetry group that makes a Virasoro algebra with a central extension \cite{Barnich}. This idea supports to existence of a two-dimensional chiral CFT as a holographic dual field theory.

One can compute the central charge of the dual quantum field theory using the manner introduced in \cite{HartmanStrominger:2009CFTDual,ChowCvetic-CFT:2008} from the NHEG (\ref{ds4d-NH}) as
\begin{equation}\label{CentarlCharge}
c=\frac{3k_i}{2\pi}\int_0^{\pi}d\theta \sqrt{F(\theta)\gamma(\theta)}
\end{equation}
Ultimately, we find the central charge associated to $2$D dual CFT 
\begin{equation}\label{Cds4}
c=\mu\frac{6l^3 r_0(l^2-r_0^2)(l^2+r_0^2)^2}{\pi(3l^6+l^4r_0^2+l^2r_0^4+3r_0^6)}.
\end{equation}
We note that, in the extremal case, the vacuum state of the bulk are in a pure state, while the dual $2$D CFT has a mixed density matrix at temperature $T_L$, is given by Eq. (\ref{kds4}) .

To ensure that there is a well defined Kerr/CFT duality, we use the Cardy's formula which gives a measure of microstates of any unitary and modular invariance $2$D CFT at a large temperature. 
\begin{equation}\label{Cardy}
S=\frac{\pi^2}{3}c_L\,T_L.
\end{equation}
So by assuming that the dual $2$D quantum field theory describes a unitary CFT, and using (\ref{Cds4}), (\ref{kds4}), the microscopic entropy of the dual CFT is computed as
\begin{equation}\label{SCFT4d}
S_{CFT}=\frac{\mu (l^2+r_0^2)}{2r_0}\sqrt{\frac{l^2-r_0^2}{2}}.
\end{equation}
that is exactly agrees to the Bekenstein-Hawking entropy of the bulk at the extremal limit (\ref{EntExt}). 

It is worth mentioning that, it was explicitly shown that for some class of charged black holes the extremality conditions and ultra-spinning limit are commutable with each other \cite{Noorbakhsh:2016}.  Moreover, It was established that the ultraspinning limit commutes with the near horizon limit. Namely, by beginning with a kerr-AdS black hole and applying both ultra-spinning and near horizon limits in different orders, one can obtained NHEG of an ultra-spinning black hole or ultra-spinning limit of a NHEG, which both results describe the same geometries. These analysis have done for general multi-spinning Kerr-AdS \cite{Mann-SEBH-CFT:2015}, and two particular classes of gauged supergravity solutions \cite{Noorbakhsh:2016}. Also for a specific class of extremal Kerr-AdS black holes whose entropy vanishes, this commutativity has been confirmed\cite{Noorbakhsh:201708}.  Now, we explore this statement for our obtained ultra-spinning charged black hole in the $4$D case  (\ref{ds4d}). 

\textbf{Extremality under Ultra-spinning limit: } We have already derived the extremality conditions for the ultra-spinning version in (\ref{Ext4d}). Now, to explore the preserving extremality under ultra-spinning limit, one need to find the extremality conditions of the origin $4$D KK-AdS black hole, which is given by
\begin{equation}\label{Ext4d0}
m=\frac{(l^2+r_0^2)(r_0^2+a^2)}{2r_0(r_0^2s^2-l^2)}, \qquad \qquad s^2=\frac{a^2(l^2-r_0^2-3r_0^4-l^2r_0^2)}{4mr_0^3}.
\end{equation}
Now, upon taking $a \rightarrow l$ limit onto these relations, they  agree with the conditions (\ref{Ext4d}). Therefore one can confirm again that extremality conditions preserve under the ultra-spinning limit.

In order to get ensured about the commutation NHEG and ultra-spinning limit with each other, we should find firstly the near horizon geometry of the $4$D KK-AdS black hole in the extremal limit, that is also presented in \cite{KKAdSEnt:Wu:2011}. Now by applying procedures towards an ultra-spinning version, one can straightforwardly check that the given NHEG exactly gets the metric (\ref{ds4d-NH}). 

It should be noted, however the main KK-AdS metric (\ref{Metric1}) and its NHEG that is given in ref. \cite{KKAdSEnt:Wu:2011} are presented in the asymptotically static frame. To find the ultra-spinning version of the NHEG, we need not to rewrite NHEG in the ARF. Namely, the near horizon geometry is free of the asymptotic behavior. So, only change coordinate (\ref{Newphi}) and taking limit $a \rightarrow l$ onto the NHEG gets the metric (\ref{ds4d-NH}). In the other words, the NHEG of an ultra-spinning black hole is equivalent to an ultra-spinning version of a NHEG of origin one. Also the sector $S^2$ of the NHEG  (\ref{ds4d-NH}) enjoys a non-compactness structure.

\subsection{Kerr/CFT description of general higher-dimensional ultraspinning black holes }
In this section, we shall investigate the Kerr/CFT correspondence for the general higher-dimensional ($D>4$) ultra-spinning KK-AdS black holes (\ref{MetricUS1}). 

To find the near horizon geometry of the metric (\ref{MetricUS1}) in the extremal case, one can find the extremality conditions by imposing $T|_{r=r_0}=0$ in (\ref{Thermo4d}), and $\hat{\Delta}|_{r=r_0}=0$. Then, by taking the following coordinate transformation 
 \begin{eqnarray}\label{NHG-Coord}
r&=&r_0(1 +\lambda  \hat{r}),\qquad  \varphi_j=\hat{\varphi_j}+\Omega_j^0 \hat{t}, \qquad
\phi_i=\hat{\phi_i}+\Omega_i^0 \hat{t}, \qquad  t=\frac{\hat{2 Y_0}}{r_0 \Delta_0^{\prime\prime} \lambda }\hat{t},\qquad \hat{\mu}_i=\mu_i.
\end{eqnarray}
and sending $\lambda \rightarrow 0$, we derived explicitly the NHEG as 
\begin{eqnarray}\label{dsKK-NH} \nonumber
ds^2&=&\hat{H}_0^{1/(D-2)}\bigg[\frac{2 \hat{U}_0}{\Delta_0^{\prime\prime}}\big(-\hat{r}^2 d\hat{t}^2 + \frac{d \hat{r}^2}{\hat{r}^2}\big) +\sum_{i,k\neq j}^{N} \tilde{g}_{ik }(d\hat{\phi}_i+k_i\,\hat{r}\,d\hat{t})(d\hat{\phi}_k+k_k\,\hat{r}\,d\hat{t})\\ 
&+&\sum_{i }^{N} g_{ij}(d\hat{\phi}_i+k_i\,\hat{r}\,d\hat{t})(d\hat{\varphi}_j+k_j\hat{r}\,d\hat{t})+ d\hat{\Omega}_0^2
\bigg],
\end{eqnarray}
where 
\begin{eqnarray}\nonumber
\tilde{g}_{ik\neq j}&=&\frac{r_0^2+a_i^2}{\Xi_i}\mu_i^2\delta_{ik}+\bigg(\frac{2m}{\hat{U}_0\hat{H}^0}+\frac{\hat{\Delta}_0[l^2s^2\hat{\Delta}_0+\hat{U}_0\hat{H}_0(l^2-s^2r_0^2)]}{\hat{H}_0\,c^2\,r_0^{2\epsilon-4}\rho_0^4\prod_{i\neq j	}^{N}(r_0^2+a_i^2)^2(c\hat{W}-\sum_{i\neq j}^{N}Z_i \Omega_i^0)}\bigg)Z_i Z_k, \\ \nonumber
\tilde{g}_{ij}&=&(r_0^2+l^2)\mu_j^2+\bigg(\frac{2m}{\hat{U}_0\hat{H}^0}+\frac{\hat{\Delta}_0[l^2s^2\hat{\Delta}_0+\hat{U}_0\hat{H}_0(l^2-s^2r_0^2)]}{\hat{H}_0\,c^2\,r_0^{2\epsilon-4}\rho_0^4\prod_{i\neq j	}^{N}(r_0^2+a_i^2)^2(c\hat{W}-\sum_{i}Z_i \Omega_j^0)}\bigg)c\,l\,Z_i \,\mu_j^2 , \\ \nonumber
Z_i&=&\frac{a_i\sqrt{\chi_i}}{\Xi_i}\mu_i^2, \qquad \qquad   \Delta_0^{\prime\prime}=\partial_{r}\hat{\Delta}|_{r=r_0}, \qquad \qquad d\hat{\Omega}_0^2=d\Omega_s^2|_{r=r_0,\mu=\hat{\mu}},
\end{eqnarray}
and 
\begin{eqnarray}\label{Ki}
k_{i\neq j}&=&-\frac{2 c\, r_0^{(\epsilon-2)}\rho_0^4\prod_{i\neq j	}^{N}(r_0^2+a_i^2)}{l\,\sqrt{l^2-s^2r_0^2}}\frac{\partial_{r_0} \Omega_i^0}{\Delta_0^{\prime\prime}},  \qquad k_{j}=-\frac{2 c\, r_0^{(\epsilon-2)}\rho_0^4\prod_{i\neq j	}^{N}(r_0^2+a_i^2)}{l\,\sqrt{l^2-s^2r_0^2}}\frac{\partial_{r_0} \Omega_j^0}{\Delta_0^{\prime\prime}}.  \end{eqnarray}

This result can be adopted in the general form of discussed in \cite{ChowCvetic-CFT:2008}
\begin{eqnarray}\label{NHGgeneral}
ds^2&=&\Gamma(y)\big(-\hat{r}^2 d\hat{t}^2 + \frac{d\hat{r}^2}{\hat{r}^2}\big)+\sum_{\alpha=1}^{n-1}F_\alpha dy_\alpha^2 + \sum_{i,j=1}^{n-1}\tilde{g}_{ij}\tilde{e}_i\tilde{e}_j. 
\end{eqnarray} 
which describes the near-horizon geometry of  D-dimensional extremal rotating black holes. 

It has been shown that by imposing consistence boundary conditions, these classes of NHEG admit $N=[(d-1)/2]$ commuting copies of the Virasoro algebra. In which, each of the Virasoro algebra is generated by the following diffeomorphisms 
\begin{equation}\label{diff}
\zeta_{(n)}^{i\neq j}=-e^{in \phi_i}\frac{\partial}{\partial_{\phi_i}}-i\,n\,r\,e^{in\phi_i}\frac{\partial}{\partial_r},  \qquad \zeta_{(n)}^j=-e^{in \varphi_j}\frac{\partial}{\partial_{\varphi_j}}-i\,n\,r\,e^{in\varphi_j}\frac{\partial}{\partial_r}, \qquad n=0,\pm1,\pm2,\dots
\end{equation}
By considering the same boundary conditions as \cite{HartmanStrominger:2009CFTDual}, These diffeomorphisms provide an asymptotic symmetry algebra.
That gives $N$ following commuting centerless Virasoro algebra
\begin{equation}\label{Valgebra}
i[\zeta_m^a,\zeta_n^a]=(m-n)\zeta_{m+n}^a.
\end{equation}
The conserved charges $Q_{\zeta^i_n}$ associated to $\zeta^i_n$ make a similar algebra with central extension. The Dirac brackets of these conserved charges are attributed to a two dimensional CFT. One can find its associated central charge by \cite{ChowCvetic-CFT:2008, Mei:2010} 
\begin{eqnarray}\label{Centralgeneral}
c_i=\frac{3\,\mu}{4\pi^2}k_i \int{d^{n-1}y_\alpha \bigg(det \tilde{g}_{ij} \prod_{\alpha=1}^{n-1} F_\alpha\bigg)^{1/2} } \int {d \phi_1 \dots d \phi_i}, \qquad i=1 \dots n-1, \quad i \neq j.
\end{eqnarray}
Where this relation corresponds to the near-horizon metric (\ref{dsKK-NH}). So, one can rewrite it as follow \cite{Mei:2010}
\begin{equation}\label{Ci}
c_{i,j}=\frac{3 \mu\,k_{i,j}}{4\pi^2} A|_{r_+=r0}. 
\end{equation}
Here $A$ denotes the area of the black hole (\ref{MetricUS1}). We refer the interested reader to \cite{Mei:2010} for similar details of calculation.

Mei in \cite{Mei:2010} introduced two general ansatzs that cover all known extremal stationary and axisymmetric black holes, in which by using the Kerr/CFT correspondence he explicitly showed that the entropy of both microscopic and macroscopic sides are exactly equivalent. Interestingly, our non-compact NHEG (\ref{NHGgeneral}) can be cast in the Mei's ansatzs as well. 

In order to use Cardy's formula as a relation between the central charge and the entropy of a two-dimensional CFT, we need to find the left- and right-moving temperature associated to $2D$ CFT. We proceed by writing the extended version of the first law for our non-compact horizon black hole as
\begin{equation}\label{FirstLawUS4d}
TdS=dM-\Omega_i dJ_i -\Omega_j dJ_i- \sum_{i}^{}{\Phi_i dQ_i} - K d\mu.
\end{equation}
where the quantity $\mu$ introduced for compactifying the new coordinate $\varphi_j$ by (\ref{period}) can be regarded as the chemical potential of the black hole. It is added to the first law by its thermodynamic conjugate $K$ \cite{HennigarKubiznakMann:2014}. In the extremal condition and using the near horizon limit \ref{NHG-Coord}, we have
\begin{equation}\label{FirstLawUS4d2}
TdS=-\big[ (\Omega_i-\Omega_i^{ex}) dJ_i + (\Omega_j-\Omega_j^{ex}) dJ_j + \sum_{i=1}^{4}{(\Phi_i-\Phi_i^{ex}) dQ_i} + (K-K^{ex}) d\mu \big].
\end{equation}

One can define the Frolov--Thorne vacuum as a vacuum state of the extremal metric. To find the temperatures associated to $2D$ CFT, we expand a scalar quantum field for the general non-extremal metric in terms of eigenstates of the asymptotic energy $E$ and angular momenta $J_i, J_j$ as 
\begin{equation}\label{FrolovVacume}
\Phi=\sum_{E,J_i,J_j,l}^{} \phi_{E J_i J_j l} \, e^{-i E \hat{t} + i J_i \hat{\phi}_i+ i J_j \hat{\varphi}_j} f_l(r,\mu_i).
\end{equation}
In the extremal condition using the near horizon limit \ref{NHG-Coord}, we have
\begin{equation}\label{FrolovVacumeNH}
 e^{-i E \hat{t} + i J_i \hat{\phi}^i+ i J_j \hat{\varphi}_j} =e^{-i n_R \hat{t} + in_L^i\hat{\phi}_i+ i n_L^j \hat{\varphi}^j },
\end{equation}
where
\begin{equation}\label{FrolovVacumeNH2}
n_L^i=J_i, \qquad n_L^j=J_j, \qquad n_R= (E-\Omega^{ex}_i J^i-\Omega^{ex}_j J_j-\Phi^{ex}Q) r_0/\lambda, 
\end{equation}
The density of states is given by $\rho=e^{S}$, where $S$ is the entropy. Using the above argument and defining the left- and right-moving Frolov--Thorne temperatures, one can write the density matrix by following Boltzmann weighting factor 
\begin{equation}\label{FrolovVacumeNH3}
e^{-(E-\Omega_iJ^i-\Omega_jJ^j-\Phi Q)/T_H}=e^{-(n_R/T_R)-(n_L/T_L)-Q/T_e}
\end{equation}
where the temperatures of the left and right-moving
CFTs are given by 
\begin{equation}\label{TF}
T_R \propto|_{ex} T_H=0, \qquad 
T_{L_{\phi_i}}=-\frac{\partial T_H/ \partial_{r_+}}{\partial \Omega_{\phi_i}/\partial_{r_+}}|_{ex}=\frac{1}{2\pi\,k_i}, \qquad \quad T_{L_{\varphi_j}}=-\frac{\partial T_H/ \partial_{r_+}}{\partial \Omega_{\varphi_j}/\partial_{r_+}}|_{ex}=\frac{1}{2\pi\,k_j}.
\end{equation}
Additionally, one can define the quantity $T_e$ associated to the electric charge as 
\begin{equation}\label{Te}
T_{e}=-\frac{\partial T_H/ \partial_{r_+}}{\partial \Phi_{\phi_i}/\partial_{r_+}}|_{r=r_0}.
\end{equation}
The important relation (\ref{TF}) was firstly used in four dimensions by \cite{HartmanStrominger:2009CFTDual}. Then it speculated to be valid for higher dimensions in \cite{ChowCvetic-CFT:2008}. It was also proved by \cite{Mei:2010} that one can generalize the definition (\ref{TF}) for all known stationary and axisymmetric extremal black holes. Moreover it was shown that for the ultra-spinning noncompactness horizon solutions, the relation (\ref{TF}) still be valid \cite{Mann-SEBH-CFT:2015}.  

The relation (\ref{TF}) has an essential role to check the validity of the Kerr/CFT correspondence. It should be noted that in $D>4$ one can find a chiral CFT arises from a Virasoro algebra corresponding to each of the $N=[D-1]/2$ rotational 2-planes. Interestingly, the central charges associated to the different CFTs which are given by equation \ref{Ci} are different, also their Frolov--Thorne temperatures differ too.

Now, helping the Cardy's formula \ref{Cardy}, which gets the entropy of an unitary CFT at temperature $T_L$, and using the relations  (\ref{Ci}) and (\ref{TF}), we find microscopic entropy of each CFTs (no summation over $i$)
\begin{equation}\label{SCFT}
S_{CFT}=\frac{\pi^2}{3}c_1T_{\phi_1}=\frac{\pi^2}{3}c_2T_{\phi_2}=\dots=\frac{\pi^2}{3}c_jT_{\varphi_j}\dots=\frac{\pi^2}{3}c_iT_{\phi_i}=\frac{A|_{r=r_0}}{4}=S_{BH}|_{ex}.
\end{equation}
This result Clearly confirm an identical microscopic entropy associated to each copy of the CFTs. Furthermore, all of them exactly agree with the Bekenstein--Hawking entropy of the ultra-spinning charged black hole at extremal limit.

Another significance feature of our ultra-spinning black holes is the contribution of the dimensionless parameter $\mu$ in all central charges and the entropy of both sides. However, this parameter is introduced for compactifying the new azimuthal coordinate $\varphi_j$ that is chosen to be an ultra-spinning direction, we see its role in all central charges associated to each CFTs aeries from other $U(1)$ direction. To have a finite central charge the quantity $\mu$ should be set by a finite value. Also one can define a new quantity in the CFT side as $T_\mu$ corresponding to the chemical potential $\mu$ \cite{Mann-SEBH-CFT:2015}. 
\begin{equation}\label{Tmu}
T_{\mu}=-\frac{\partial T_H/ \partial_r{r_+}}{\partial K/\partial_r{r_+}}|_{r=r_0},
\end{equation}
Where $K$ is the conjugate variable to the chemical
potential $\mu$, introduced firstly in \cite{HennigarKubiznakMann:2014}. 

We also note, it is discussed in \cite{HartmanStrominger:2009} that, within the Einstein-Hilbert theories the non-gravitational fields have no contribution to the central charge of the dual CFTs. This argument is observed for our new ultra-spinning solutions as well. 

We note, the Mei's ansatzs in \cite{Mei:2010} which gives a well-defined Kerr/CFT correspondence, cover the most stationary and axisymmetric black holes with compact horizon. Also in \cite{KKAdSEnt:Wu:2011} was shown that the general charged KK-AdS black holes (\ref{Metric1}) can be followed by Mei's ansatzs, indicating the Kerr/CFT correspondence is applicable for this general solution as well. Now, we emphasis that for our general ultra-spinning KK-AdS black holes solution (\ref{MetricUS1}), despite the noncompactness of their horizons, the result (\ref{SCFT}) shows a precise holographically agreement between the microscopic entropy of the CFT side and the entropy of the black hole. Namely, our new class of charged black hole \ref{MetricUS1} solution exhibit a well-defined Kerr/CFT correspondence.

\section{Summery}\label{Discussion}
In this work, we have employed the ultra-spinning (super-entropic) limit as a fairly simple generating solution technique to construct a novel class of charged multi-spinning black holes as solutions of Einstein-Maxwell-Dilaton-Lambda (EMD$\Lambda$) theory in all higher dimensions $D \ge 4$. We start from a known black hole solution which refer to KK-AdS black holes, describing a higher dimensional single charged multi-spinning solution of EMD$\Lambda$ theory. Our main motivation focus on more explore to the charged black holes at ultra-spinning limit or in large angular momentum. In this regard interestingly, we succeed to generate a novel class of charged rotating black holes as exact solutions of the same theory given by action (\ref{ActionKK}). Of course, with different horizon and different conformal boundary structurer comparing to origin KK-AdS black hole.

Our constructing geometry describes an asymptotic AdS black hole with noncompact horizon in all dimensions. The topology of the event horizon present a sphere with some punctures where arises from the poles of the polar coordinate components. In four dimensional case, we explicitly have shown that these poles are located at $\theta=0, \pi$. In the higher dimensions the number of punctures are as many number as the roots of $\mu_j=0$. We emphasize that these poles can be interpreted as a sort of boundary, and they are removed from the spactime indeed. Therefore the obtained solution in all dimensions has a regular horizon. However, interestingly their horizon area and entropy are finite. 

We note that, it is impossible to generate a multi ultra-spinning solution by performing more than one rotation parameter as ultra-spinning direction. It also should be emphasized that the thermodynamic quantities cannot be obtained by taking the simple limit $a \rightarrow l$ onto the origin KK-AdS black hole quantities, because they obviously will diverge. Indeed there is no trivial relation between the thermodynamic quantities of both origin and ultra-spinning solutions. But due to compactifying the new azimuthal coordinate it is expected the conserved charges and thermodynamic quantities associated to our ultra-spinning black hole would be finite. A further research will be computing these conserved charges and investigating its thermodynamic in the extended phases space thermodynamics background to ensure whether one can find a range of parameter space leading to super-entropic black holes, similar analyze for general Kerr-AdS black hole was done in \cite{HennigarKubiznakMann:2015}.

We have also obtained the near horizon geometry of our ultra-spinning solutions in the extremal limit. That presented a direct product of an AdS$_2$ sector and a $S^{d-2}$ manifold having same punctures as origin black hole. Moreover, for four dimensional case we have explicitly investigated the Kerr/CFT correspondence, by calculating the central charge and the microscopic entropy of dual $2$D CFT. Then via the Cardy's formula we have shown that our result indicate a precise agreement between the microscopic and macroscopic entropies. 

For the general higher dimensional case, we have explicitly presented the near horizon geometry of extremal solution. In the $D$-dimensional case we are dealing with $[(D-1)/2]$ independent $U(1)$ rotation symmetries that by imposing a appropriate boundary condition one can get $[(D-1)/2]$ copies of Virasoro algebra, supporting the same number $2D$ dual CFTs. We have given a main result for their different central charges $c_i$ as well as their associated Frolov--Thorne temperatures. Also our result via the Cardy's formula  confirm an identical microscopic entropy associated to each of the CFTs, and a fully agreement with the Bekenstein--Hawking entropy of the ultra-spinning charged black hole at extremal limit. Namely, our ultra-spinning solution exhibit a well-defined Kerr/CFT correspondence in all dimensions despite its noncompact horizon.

\section*{Acknowledgments}
SMN would like to thanks K. Hajian, M. M. Sheikh-Jabbari and M. H. Vahidinaia for useful discussions on the related topics. Also SMN thanks the organizers of the “Recent Trends in String Theory and Related Topics” workshop, held in May 2017 in Tehran, at which some results of this work were presented. Also thanks the Institute for Research in Fundamental Sciences (IPM) for hospitality while this project was accomplished.

%


\begin{thebibliography}{99}%
	
\bibitem{MyersPerry:1986}
R. Myers and M. Perry, Black Holes in Higher
Dimensional Space-Times, Annals Phys. 172 (1986)
304.	
	

\bibitem{EmparanMyers:2003}
R. Emparan and R. C. Myers, Instability of ultra-spinning black holes, JHEP 0309 (2003) 025, [arXiv:hep-th/0308056].

\bibitem{CaldarelliEmparan:2008}
M. M. Caldarelli, R. Emparan, and M. J. Rodriguez, Black Rings in (Anti)-deSitter space, JHEP 0811	(2008) 011, [arXiv:hep-th/0806.1954].

\bibitem{ArmasObers}
J. Armas and N. Obers, Blackfolds in (Anti)-de Sitter Backgrounds, Phys.Rev. D83 (2011) 084039,	 [arXiv:hep-th/1012.5081].


\bibitem{CaldarelliLeigh:2011}
M. M. Caldarelli, R. G. Leigh, A. C. Petkou, P. M. Petropoulos, V. Pozzoli, et al., " Vorticity in holographic fluids, PoS CORFU2011 (2011) 076, [arXiv:hep-th/1206.4351].

\bibitem{HennigarKubiznakMann:2014}	
R. A. Hennigar, D. Kubiznak and R. B. Mann, “Super-Entropic Black Holes,” Phys. Rev. Lett. 115, no. 3, 031101 (2015), [arXiv:hep-th/1411.4309].


\bibitem{GneccchiHristovKlemm:2014}	
A. Gnecchi, K. Hristov, D. Klemm, C. Toldo, and O. Vaughan, "Rotating black holes in 4d gauged supergravity, JHEP 1401 (2014) 127, [arXiv:hep-th/1311.1795].


\bibitem{Klemm:2014}
D. Klemm, Four-dimensional black holes with unusual horizons, Phys.Rev. D89 (2014) 084007, [arXiv:hep-th/1401.3107].


\bibitem{HennigarKubiznakMann:2015}	
R. A. Hennigar, D. Kubiznak, R. B. Mann and N. Musoke, “Ultraspinning limits and super-entropic black holes,” JHEP 1506, 096 (2015), [arXiv:hep-th/1504.07529].


\bibitem{Noorbakhsh:2016}
S.M. Noorbakhsh, and M. Ghominejad,"Ultra-spinning gauged supergravity black holes and their Kerr/CFT correspondence", Phys. Rev. D 95, 046002 (2017), [arXiv:1611.02324]. 


\bibitem{Noorbakhsh:201708} 
S.~M.~Noorbakhsh and M.~H.~Vahidinia,
``Extremal Vanishing Horizon Kerr-AdS Black Holes at Ultraspinning Limit,'' arXiv:1708.08654 [hep-th].

\bibitem{KastorRayTraschen:2009}
D. Kastor, S. Ray, and J. Traschen, "Enthalpy and the Mechanics of AdS Black Holes", Class.Quant.Grav. 26
(2009) 195011, [arXiv:0904.2765].


\bibitem{CveticGibbonsKubiznak-ISO:2012}
M. Cvetic, G. Gibbons, D. Kubiznak, and C. Pope, "Black Hole Enthalpy and an Entropy Inequality for the	Thermodynamic Volume", Phys.Rev. D84 (2011) 024037, [arXiv:hep-th/1012.2888].


\bibitem{NH1}
J. W. York, Phys. Rev. D28 (1983) 2929. 
W. H. Zurek and  K. S. Thorne, Phys. Rev. Lett. 54 (1985) 2171. 
J. A. Wheeler, A Journey into Gravity and Spacetime Freeman, N.Y. (1990). 
G. ’t Hooft, Nucl. Phys. B335 (1990) 138. 
L. Susskind, L. Thorlacius and R. Uglum, Phys. Rev. D48 (1993) 3743. 
V. Frolov and I. Novikov, Phys. Rev. D48 (1993) 4545. 
S. Carlip, Phys. Rev. D51 (1995) 632. 
M. Cvetic and A. Tseytlin, Phys. Rev. D53 (1996) 5619. 
F. Larsen and F. Wilczek, Phys. Lett. B375 (1996) 37.
\bibitem{NH10}
A. Strominger. Black hole entropy from near horizon microstates. JHEP, 02:009, 1998.

\bibitem{NH11} 
S. W. Hawking, M. J. Perry, and A. Stro-
minger, Phys. Rev. Lett. 116, 231301 (2016), arXiv:1601.00921 [hep-th] .

\bibitem{NH12} 
H. Afshar, D. Grumiller and M. M. Sheikh-Jabbari, Near Horizon Soft Hairs as Microstates of Three Dimensional Black Holes," [arXiv:hep-th/1607.00009].

\bibitem{GuicaHartmanStrominger-KerrCFT}		
M. Guica, T. Hartman, W. Song and A. Strominger, "The Kerr/CFT correspondence," [arXiv:hep-th/0809.4266].

\bibitem{BrownHenneaux:1986}
J. D. Brown and M. Henneaux, Comm. Math. Phys. 104 (1986) 207.

\bibitem{Cardy}
J. Cardy, “Operator Content Of Two-Dimensional Conformally-Invariant Theories,” Nucl. Phys. B270 (1986) 186.

\bibitem{HartmanStrominger:2009CFTDual}	
T. Hartman, K. Murata, T. Nishioka and A. Strominger, “CFT Duals for Extreme Black Holes”, JHEP 0904 (2009) 019 [arXiv:0811.4393 [hep-th]].
	
\bibitem{LuMeiPop-CFT:2008}	
H. Lu, J. Mei and C.N. Pope, Kerr–AdS/CFT correspondence in diverse dimensionarXiv:0811.2225 [hep-th].
	
\bibitem{ChowCvetic-CFT:2008}
D. Chow, M. Cvetic, H. Lu, and C. Pope, ”Extremal Black Hole/CFT Correspondence in (Gauged) Supergravities,” [arXiv:hep-th/0812.2918v2].	

\bibitem{Compere:2009}
G. Compere, K. Murata and T. Nishioka, Central Charges in Extreme Black Hole/CFT
Correspondence, JHEP 0905, 077 (2009) [arXiv:0902.1001 [hep-th]]. 
O. J. C. Dias, H. S. Reall and J. E. Santos, “Kerr-CFT and gravitational perturbations,” JHEP 0908 (2009) 101, [arXiv:0906.2380 [hep-th]].


\bibitem{Mann-SEBH-CFT:2015}
M. Sinamuli and R. B. Mann, "Super-Entropic Black Holes and the Kerr-CFT Correspondence", [arXiv:hep-th/1512.07597].


\bibitem{KK4d:1987}
V. Frolov, A. Zelnikov, and U. Bleyer, Ann. Phys.
(Leipzig) 499, 371 (1987).

\bibitem{KK4dNUT:2008}
A.N. Aliev, H. Cebeci, and T. Dereli, Phys. Rev. D 77, 124022 (2008).

\bibitem{KKd:2006}	
J. Kunz, D. Maison, F. Navarro-Lerida, and J. Viebahn,
Phys. Lett. B639, 95 (2006).

\bibitem{KKgauged1:2005}
Z.W. Chong, M. Cvetiˇc, H. L¨u, and C.N. Pope, Phys.
Rev. D 72, 041901 (2005).

\bibitem{KKgauged2:2005}
D.D.K. Chow, Classical Quantum Gravity 28, 032001 (2011).


\bibitem{KK-AdS:Wu:2011}
S.-Q. Wu, General rotating charged Kaluza-Klein-AdS black holes in higher dimensions, Phys. Rev. D 83 (Jun, 2011) 121502. [arXiv:hep-th/1108.4157v1] 



\bibitem{Gibbons:KerrAdS1}		
G. W. Gibbons, H. Lu, D. N. Page, and C. N. Pope,
Rotating black holes in higher dimensions with a
cosmological constant, Phys. Rev. Lett. 93 (2004)
171102, [hep-th/0409155].

\bibitem{Gibbons:KerrAdS2}		
G. W. Gibbons, H. Lu, D. N. Page, and C. N. Pope, The general Kerr-de Sitter metrics in all dimensions, J. Geom. Phys. 53 (2005) 49-73, [hep-th/0404008].	


\bibitem{Kunz:2006}		
J. Kunz, D. Maison, F. Navarro-Lerida and J. Viebahn,  Rotating Einstein-Maxwell-Dilaton Black Holes in D Dimensions, Phys. Lett. B 639, 95 (2006)
[arXiv:hep-th/0606005].

\bibitem{Cvetic:1999}		
M. Cvetic, M. J. Duff, P. Hoxha, J. T. Liu, H. Lu, J. X. Lu, R. Martinez-Acosta, C. N. Pope et al., “Embedding AdS black holes in ten-dimensions and elevendimensions,” Nucl. Phys. B558 (1999) 96-126. [hep-th/9903214].


\bibitem{KunduriLucietti}
H. K. Kunduri, J. Lucietti and H. S. Reall, “Near-horizon symmetries of extremal black holes,” Class. Quant. Grav. 24 (2007) 4169, [arXiv:0705.4214 [hep-th]]. 

\bibitem{KunduriLucietti2}
H. K. Kunduri and J. Lucietti, “A classification of near-horizon geometries of extremal vacuum black holes,” [arXiv:0806.2051 [hep-th]]. H. K. Kunduri and J. Lucietti, “Classification of near-horizon geometries of extremal	black holes,”Living Rev. Rel. 16 (2013) 8, [arXiv:1306.2517 [hep-th]].
	
	
\bibitem{Astefanesei}	
D. Astefanesei, K. Goldstein, R. P. Jena, A. Sen, and S. P. Trivedi, Rotating attractors, JHEP
0610:058,2006 (JHEP 0610:058,2006) [hep-th/0606244].

\bibitem{JohnstoneSheikh-Jabbari}
M. Johnstone, M.M. Sheikh-Jabbari, J. Simon and H. Yavartanoo, “Extremal Black	Holes and First Law of Thermodynamics,” [arXiv:1305.3157 [hep-th]].	

\bibitem{Barnich}	
G. Barnich and F. Brandt, “Covariant theory of asymptotic symmetries, conservation laws and central charges,” Nucl. Phys. B 633, 3 (2002) [arXiv:hep-th/0111246]. G. Barnich and G. Compere, “Conserved charges and thermodynamics of the spinning Goedel black hole,” Phys. Rev. Lett. 95, 031302 (2005) [arXiv:hep-th/0501102]. 

\bibitem{Mei:2010}					
Mei, J., “The Entropy for General Extremal Black Holes”, J. High Energy Phys., 2010(04),
(2010). [DOI], [arXiv:1002.1349 [hep-th]].

\bibitem{FrolovThorne}	
V.P. Frolov and K.S. Thorne, Renormalized stress-energy tensor near the horizon of a slowly evolving, rotating black hole, Phys. Rev. D39, 2125 (1989).


\bibitem{KKAdSEnt:Wu:2011}		
J. -J. Peng, S.-Q. Wu, Statistical entropies of extremal Kaluza-Klein AdS black holes in arbitrary dimensions, Gen Relativ Gravit (2012) 44:993–1005. 		

\bibitem{HartmanStrominger:2009}	
T. Hartman, K. Murata, T. Nishioka and A. Strominger, CFT Duals for Extreme Black
Holes, JHEP 0904, 019 (2009) [arXiv:0811.4393 [hep-th]].
	
\end{thebibliography}
\end{document}